\documentclass[twocolumn,prc,aps,showpacs,nofootinbib,amsmath,amssymb]{revtex4}
\usepackage{epsfig}
\usepackage{graphicx}% Include figure files
\usepackage{dcolumn}
\usepackage{bm}

\usepackage{pslatex}

 \newcommand\beq{\begin{equation}}
 \newcommand\eeq{\end{equation}}
 \newcommand\beqn{\begin{eqnarray}}
 \newcommand\eeqn{\end{eqnarray}}

\def\GeV{\,\mbox{GeV}}

\def\GeV{\,\mbox{GeV}}

\def\eq#1{{Eq.~(\ref{#1})}}

\begin{document}

\title{{\Large Direct photons at forward rapidities in high-energy pp collisions }}

\author{B.~Z.~Kopeliovich$^{1,2}$}
\author{E.~Levin$^3$}
\author{A.~H.~Rezaeian$^1$}
\author{Ivan~Schmidt$^1$}
 
 \affiliation{$^1$Departamento de F\'\i sica y Centro de Estudios
Subat\'omicos,\\ Universidad T\'ecnica
Federico Santa Mar\'\i a, Casilla 110-V, Valpara\'\i so, Chile\\
{$^2$Joint Institute for Nuclear Research, Dubna, Russia}\\
 {$^3$Department of Particle Physics, School of Physics and Astronomy,
Raymond and Beverly Sackler Faculty of Exact Science, Tel Aviv University, Tel Aviv, 69978, Israel }}

\begin{abstract}
We investigate direct photon production in $pp$ collisions at the
energies of RHIC, CDF and LHC, at different rapidities employing
various color-dipole models. The cross section peaks at forward
rapidities due to the abelian dynamics of photon radiation. This opens
new opportunities for measurement of direct photons at forward
rapidities, where the background from radiative hadronic decays is
strongly suppressed. Our model calculations show that photon
production is sensitive to the gluon saturation effects, and strongly
depends on the value of the anomalous dimension.
\end{abstract}

\maketitle

\date{\today}
\section{Introduction}

Photons radiated in hadronic collisions not via hadronic decays are
usually called direct. They carry important information about the
collision dynamics, not disturbed by final state interactions. In
particular, the hadronization stage is absent, so the theoretical
interpretation is simpler that in the case of hadron production. A
unified description for radiation of virtual (Drell-Yan reaction) and
real photons within the color dipole approach was proposed in
\cite{hir,kst1}.  This description does not need to be corrected
either for higher order effects (K-factor, large primordial transverse
momentum), or for the main higher twist terms\footnote{Some higher
twist corrections, which are specific for forward rapidities, still
have to be added \cite{progress}. }.  The corresponding phenomenology
is based on the universal dipole cross section fitted to DIS data and
provides a rather good description of data, both the absolute
normalization and the transverse momentum dependence
\cite{amir0}. Predictions of the inclusive direct photon spectra for the LHC at midrapidity 
and the azimuthal asymmetry of produced prompt photons in the same
framework are given in Refs.~\cite{us-3,asy0}. Comparison with the
predictions of other approaches at the LHC can be found in
Refs.~\cite{lhc-hic,nlhc}.

Intensive study of the dynamics of hadronic interactions at high energies and
search for signatures of nonlinear QCD effects, like saturation \cite{glr}, or color glass condensate (CGC)
\cite{mv}, have led to considerable experimental progress towards reaching smallest Bjorken $x$.
The typical experimental set up at modern colliders allows to detect
particles produced in hard reactions in the central rapidity region,
while the most energetic ones produced at forward (backward)
rapidities escape detection. The first dedicated measurements of
hadron production at forward rapidities, by the BRAHMS experiment in
deuteron-gold collisions at RHIC \cite{brahms}, disclosed an
interesting effect of nuclear suppression, which can be interpreted as
a breakdown of QCD factorization \cite{knpjs}.  The zero degree
calorimeters detecting neutral particles, neutrons and photons, at
maximal rapidities are being employed at RHIC and are planned to be
installed at LHC. One could increase the transverse momentum coverage
of these detectors by moving them away from the beam axis. Through this note
we would like to encourage the experimentalists to look up at these
opportunities.

The important advantage of measurements of direct photons at forward
rapidities is a significant enhancement of the signal-to-background
ratio. Indeed, the photons radiated by the electric current of the
projectile quarks, which stay in the fragmentation region of the beam,
form a bump at forward rapidities (see Fig.~5). At the same time,
gluons are radiated via nonabelian mechanisms by the color current
which flows across the whole rapidity interval. Therefore gluons are
radiated mainly in the central rapidity region \cite{gb}, and are
strongly suppressed in the beam fragmentation region. Such a
suppression is even more pronounced for hadrons from gluon
fragmentation, and for photons from radiative decays of those
hadrons. Another source of background photons, hadrons produced via
fragmentation of the valence quarks is also strongly suppressed due to
the shift towards small fractional momenta related to the quark
fragmentation function and the kinematics of radiative decay. Thus,
direct photon production at forward rapidities should be substantially
cleared up.

Here we perform calculations for direct photon production at large $p_T$ and various rapidities in proton-proton collisions at the energies of RHIC, Tevatron and LHC. We employ the color dipole approach and compare the predictions of several contemporary models for the dipole cross section, based on the idea of gluon saturation.

\section{Photon radiation in the color dipole formalism} 

\begin{figure}[!t]
       \centerline{\includegraphics[width=6 cm] {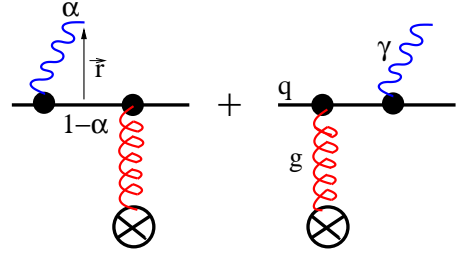}} \caption{
       Direct photon radiation by a quark interacting with a gluon in the target.}
      \label{fr1}
       \end{figure}

Production of direct photons in the target rest frame should be treated as
electromagnetic bremsstrahlung by a quark interacting with the target, as is illustrated in Fig.~\ref{fr1} in the single gluon approximation, which should be accurate at large transverse momenta of the photon.
Since the quark trajectories before and after photon radiation have different impact parameters,
and the corresponding terms in the bremsstrahlung amplitude have different signs, one arrives at an expression, which is formally identical to the amplitude of an inelastic dipole-target interaction \cite{hir}.
This is only a formal procedure of calculation, while no real dipole is involved in the process of radiation.  

Calculation of the transverse momentum distribution is more involved \cite{kst1}, since the direct and complex conjugated amplitudes correspond to incoming quarks with different impact parameters. It substantially simplifies after integration over transverse momentum of the recoil quark, so one is left with two dipoles of different sizes $\vec r_1$ and $\vec r_2$, and the cross section gets the factorized form \cite{kst1,amir0},
 
\begin{eqnarray}
&&\frac{d \sigma(qp\to\gamma X)}
{d(ln \alpha)\,d^{2}{\vec{p}_T}}(\vec{p}_T,\alpha)=\frac{1}{(2\pi)^{2}}
\sum_{in,f}\int d^{2}{r}_{1}d^{2}{r}_{2}\nonumber\\
&\times&
e^{i \vec{p}_T\cdot
(\vec{r}_{1}-\vec{r}_{2})}
\phi^{\star}_{\gamma q}(\alpha, \vec{r}_{1})
\phi_{\gamma q}(\alpha, \vec{r}_{2})
\Sigma_{\gamma}(x_2,\vec{r}_{1},\vec{r}_{2},\alpha), \label{m1}
\end{eqnarray} 
where $\vec{r}_{1}$ and $\vec{r}_{2}$ are the quark-photon transverse
 separations in the direct and complex conjugated amplitudes respectively;
 $\alpha=p_\gamma^+/p_q^+$ denotes the fractional light-cone (LC) momentum of the radiated
photon.  Correspondingly, the transverse displacements of the recoil quarks in the two amplitudes are $\alpha r_{1}$ and $\alpha r_{2}$ respectively. The LC
distribution amplitude for the $q\gamma$ Fock component with
transverse separation $\vec r$ has the form, 
\beq
\phi_{\gamma q}(\alpha,\vec r_T)=
\frac{\sqrt{\alpha_{em}}}{2\,\pi}\,
\chi_f\,\widehat O\,\chi_i\,K_0(m_q \alpha r_T).
\label{dylcl}
\eeq
Here $\chi_{i,f}$ are the spinors of the initial and final quarks and
$K_0(x)$ is the modified Bessel function.  The operator $\widehat O$
has the form,
\beq
\widehat O = i\,m_f\alpha^2\,
\vec {e^*}\cdot (\vec n\times\vec\sigma)\,
 + \alpha\,\vec {e^*}\cdot (\vec\sigma\times\vec\nabla)
-i(2-\alpha)\,\vec {e^*}\cdot \vec\nabla\ ,
\eeq
where $\vec e$ is the polarization vector of the photon; $\vec n$ is a
unit vector along the projectile momentum; and $\vec\nabla$ acts on
$\vec r_T$. The effective quark mass $m_q$  serves as
infra-red cutoff parameter, which we fix at  $m_q\approx 0.14 \GeV$, since all dipole parametrizations considered in this paper also assume the light quark mass equal to $0.14$ GeV. 

In equation (\ref{m1}) the effective dipole cross-section $\Sigma_{\gamma}$ is a linear combination of the $\bar qq$ dipole-proton cross sections,
 \begin{eqnarray}
\Sigma_{\gamma}(x_2,\vec{r}_{1},\vec{r}_{2},\alpha)&=&\frac{1}{2}\{
\sigma_{q\bar{q}}(x_2,\alpha r_{1})+\sigma_{q\bar{q}}(x_2,\alpha r_{2})\}\nonumber\\
&-&\frac{1}{2}\sigma_{q\bar{q}}(x_2,\alpha(\vec{r}_{1}-\vec{r}_{2})).
\label{sig}
\end{eqnarray}
Here and in what follows $x_{1,2}$ are the  Bjorken variable of the beam and target partons,
\beq
x_{1,2}=\frac{p_T}{\sqrt{s}}\,e^{\pm \eta},
\label{x2}
\eeq
and $\eta$ is the photon rapidity in the c.m. of $pp$ collision.

Since only quarks and antiquarks can radiate photons, the hadronic cross section is given by the convolution of the partonic cross section Eq.~(\ref{m1}) with the proton structure function 
$F_{2}^{p}$ \cite{joerg,amir0,scho},
\begin{equation}
\frac{d\sigma(pp\to\gamma X)}{dx_{F}d^{2}\,{p}_{T}}=
\frac{1}{x_1+x_2}
\int\limits_{x_1}^1 d\alpha \, F_{2}^{p}\left({x_1\over\alpha},Q^2\right)\, 
\frac{d \sigma(qp\to \gamma X)}{d\alpha\,d^{2}{p}_{T}},
 \label{con1}
\end{equation}
where $x_F=x_1-x_2$ is the Feynman variable. 
 This relation needs commenting. The transverse momentum distribution of quark bremsstrahlung should be convoluted with the primordial transverse motion of the projectile quark. Differently from the parton model, in the dipole approach one should rely on the quark distribution function taken at a soft scale.
Evolution to the hard scale is performed via gluon radiation, which is encoded into the phenomenological dipole cross section fitted to DIS data for the proton structure function.  Since the quark primordial motion with a small (soft) mean transverse momentum does not affect the photons radiated with large $p_T$ \cite{amir0}, we neglect the transverse momentum convolution and use the integrated quark distribution.

However, a word of caution is in order. The dipole cross section $\sigma_{\bar qq}(x_2,r)$ includes gluon radiation which performs the $Q^2$ evolution and leads to an increase of the  transverse momentum of the projectile quark. However, it misses the $Q^2$ evolution of the $x_1$-distribution, which is especially important at forward rapidities (see discussion in \cite{bgks}), since the quark distribution falls off at $x_1\to1$ much steeper at high $Q^2$. In order to account for this effect and provide the correct $x_1$-distribution, we take the integrated quark distribution in (\ref{con1}) at the hard scale.

We use the parametrization
for the proton structure function given in Ref.~\cite{ps}. Following
Ref.~\cite{amir0} , 
for the hard scale entering the proton structure function in
Eq.~(\ref{con1}), we choose $Q=p_T$.

\section{Models for the dipole cross section}

The dipole cross-section is theoretically unknown and should be fitted to data.  Several
parametrizations proposed in the literature are employed here to investigate the
uncertainties and differences among various models.
 
A popular parametrization proposed by Golec-Biernat and W\"usthoff (GBW)
model \cite{gbw} is based on the idea of gluon saturation. This model is able to describe
DIS data with the dipole cross-section parametrized as, 
\begin{equation}
\sigma_{q\bar{q}}^{\text{GBW}}(x,\vec{r})=\sigma_{0}\left(1-e^{-r^{2}Q_{s}^{2}(x)/4}\right), \label{gbw}
\end{equation}
where $x$-dependence of the saturation scale is given by $Q_s^2(x)=(x_0/x)^{\lambda}~\text{GeV}^2 $.
The parameters $\sigma_{0}=20.1$ mb, $x_{0}=5.16\times 10^{-4}$, and $\lambda=0.289$ were determined from a fit to $F_2$ without
charm quarks. A salient feature of the model is that for decreasing $x$, the
dipole cross section saturates at smaller dipole sizes. 
The saturation scale in the GBW reduces with the inclusion of the charm quark \cite{gbw-nn}.   
After inclusion of the charm quark with mass $m_c=1.5$ GeV, 
the parameters of the GBW model changed to $\sigma_{0}=23.9$ mb,
$x_{0}=1.11\times 10^{-4}$ , and $\lambda=0.287$. Both parametrization sets give a good description of DIS data at
$x<0.01$ and $Q^2\in[0.25,45]$ \cite{gbw-nn}. 
\begin{figure}[!t]
       \centerline{\includegraphics[width=8 cm] {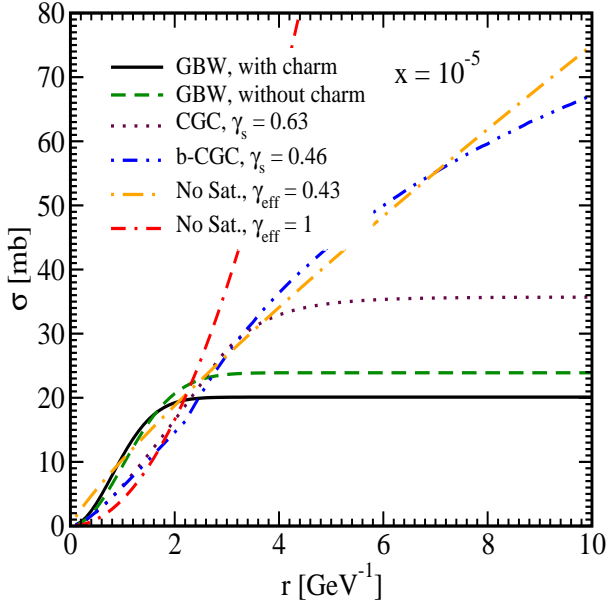}} \caption{
       The total dipole-proton cross section $\sigma_{q\bar{q}}(x, r)$ at a fixed $x= 10^{-5}$
        in various dipole models introduced in Sec. III.
       \label{fd} }
\end{figure}

One of the obvious shortcoming of  the GBW model is
that it does not match the QCD evolution (DGLAP) at large values of
$Q^{2}$. This failure can be clearly seen in the energy dependence
of $\sigma^{\gamma^{\star} p}_{tot}$ for $Q^{2}> 20~\text{GeV}^{2}$, where the
the model predictions are below the data \cite{gbw,gbw-d}.

A modification of the GWB dipole parametrization model,
Eq.~(\ref{gbw}), was proposed in Ref.~\cite{gbw-d} by Bartels, Golec-Biernat and Kowalski (GBW-DGLAP)
\begin{equation}
\sigma_{q\bar{q}}^{\text{GBW-DGLAP}}(x,\vec{r})=\sigma_{0}\left(1-exp\left(-\frac{\pi^{2}r^{2}
\alpha_{s}(\mu^{2})xg(x,\mu^{2})}{3\sigma_{0}}\right)\right), \label{gbw1}
\end{equation}
where the scale $\mu^{2}$ is related to the dipole size by
\begin{equation}
\mu^{2}=\frac{C}{r^{2}}+\mu_{0}^{2}. \label{scale}
\end{equation}
Here the gluon density $g(x,\mu^{2})$ is evolved to the scale
$\mu^{2}$ with the leading order (LO) DGLAP equation.  Moreover, the
quark contribution to the gluon density is neglected in the small $x$
limit. The initial gluon density is taken at the scale $Q_{0}^{2}=1
\text{GeV}^{2}$ in the form
\begin{equation} xg(x,\mu^{2})=A_{g}x^{-\lambda_{g}}(1-x)^{5.6},
\end{equation}
where the parameters $C=0.26$, $\sigma_0=23~\text{mb}, \mu_{0}^{2}=0.52~\text{GeV}^{2}$,
$A_{g}=1.20$ and $\lambda_{g}=0.28$ are fixed from a fit to the DIS
data for $x<0.01$ and in a range of $Q^2\in[0.1,500]$
$\text{GeV}^2$ \cite{gbw-d}. The dipole size determines
the evolution scale $\mu^{2}$ through Eq.~(\ref{scale}). The
evolution of the gluon density is performed numerically for every
dipole size $r$ during the integration of Eq.~(\ref{m1}). Therefore,
the DGLAP equation is now coupled to our master equation
(\ref{m1}). It is important to stress that the GBW-DGLAP
model preserves the successes of the GBW model at low $Q^{2}$ and
its saturation property for large dipole sizes, while incorporating
evolution of the gluon density by modifying the small-$r$
behaviour of the dipole size.

Since the linear DGLAP evolution may not be appropriate for the saturation regime,
Iancu, Itakura and Munier proposed
an alternative color glass condensate (CGC)
model \cite{CGC0}, based on the Balitsky-Kovchegov (BK) equation \cite{bk}.
The dipole cross section is parametrized as, 
\begin{equation} \label{cgc}
  \sigma_{q\bar{q}}^{\text{CGC}}(x,r) =\sigma_0
  \begin{cases}
    \mathcal{N}_0\left(\frac{rQ_s}{2}\right)^{2\left(\gamma_s+
    \frac{1}{\kappa\lambda Y}\ln\frac{2}{rQ_s}\right)} & :\quad rQ_s\le 2\\
    1-\mathrm{e}^{-A\ln^2(BrQ_s)} & :\quad rQ_s>2
  \end{cases},
\end{equation}
where $Q_s\equiv Q_s(x)=(x_0/x)^{\lambda/2}$ GeV, $Y=\ln(1/x)$, and $\kappa = \chi''(\gamma_s)/\chi'(\gamma_s)$ where $\chi$ is the LO BFKL characteristic function.  
The coefficients $A$ and $B$ in the second line of \eqref{cgc} are determined uniquely from the condition that $\sigma(x,r)$, 
and its derivative with respect to $rQ_s$, are continuous at $rQ_s=2$:
\begin{equation} \label{eq:AandB}
  A = -\frac{\mathcal{N}_0^2\gamma_s^2}{(1-\mathcal{N}_0)^2\ln(1-\mathcal{N}_0)}, \qquad B = \frac{1}{2}\left(1-\mathcal{N}_0\right)^{-\frac{(1-\mathcal{N}_0)}{\mathcal{N}_0\gamma_s}}.
\end{equation}
The parameters $\gamma_s=0.63$ and $\kappa=9.9$ are fixed at the LO
BFKL values. The others parameters $\mathcal{N}_0=0.7$,
$\sigma_0=35.7$ mb, $x_0=2.7\times 10^{-7}$ and $\lambda=0.177$ were
fitted to $F_2$ for $x<0.01$ and $Q^2<45$ $\text{GeV}^2$ and including
a charm quark with $m_c=1.4$ GeV. Notice that for small $rQ_s\le 2$,
the effective anomalous dimension $1-\gamma_s$ in the exponent in the
upper line of Eq.~(\ref{cgc}) rises from the LO BFKL value towards the
DGLAP value. 

It should be stressed that this CGC model is built based on the solution of Ref.~\cite{LT} for $r\,Q_s \,>2\,$ and 
a form of the solution for $r Q_s \leq 1$, but in the vicinity of  $r \propto 1/Q_s$ it is given in Refs.~\cite{IIM,MUT}.

\begin{figure}[!t] 
\centerline{\includegraphics[width=8
cm] {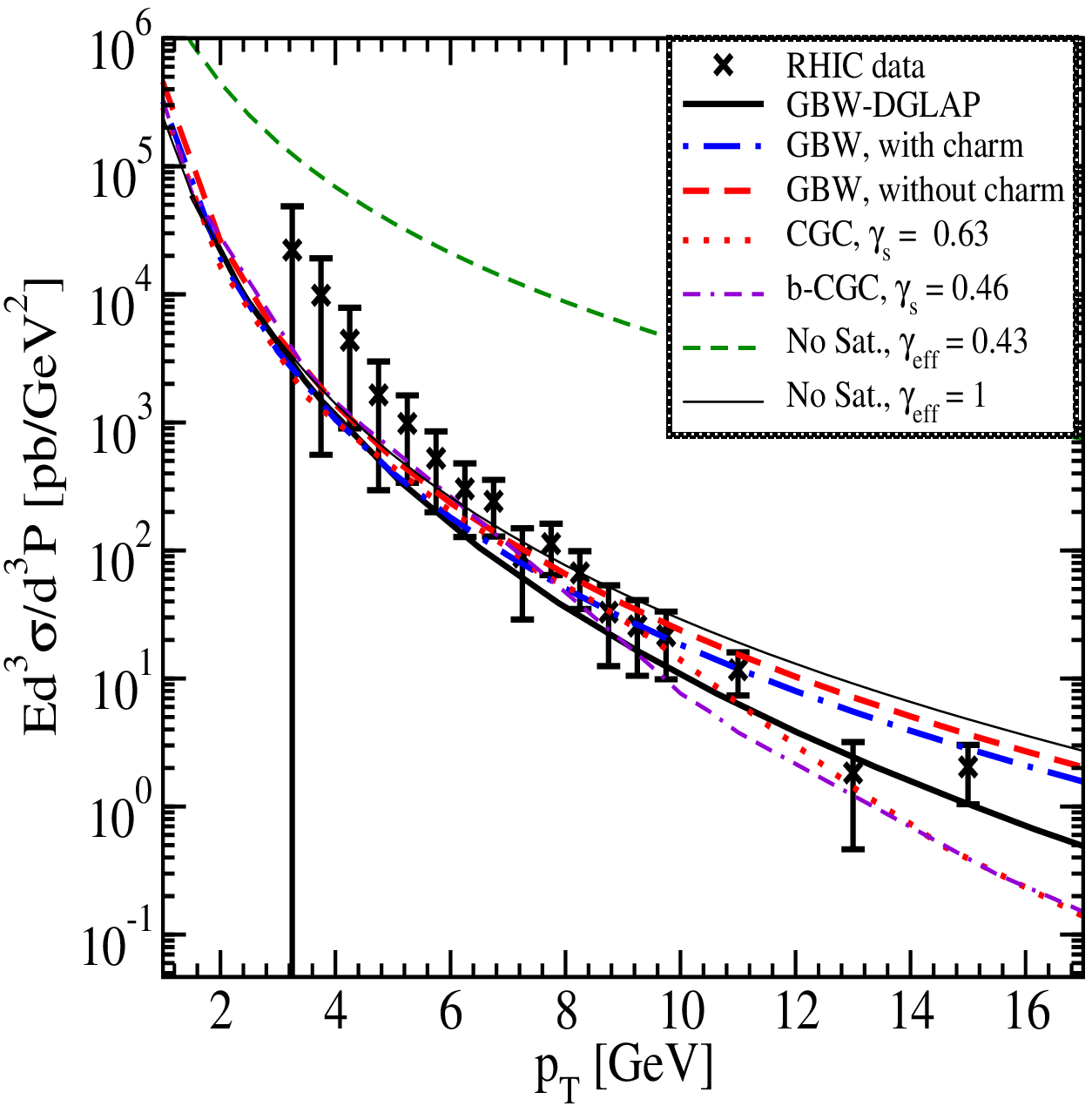}} \centerline{\includegraphics[width=8 cm]
{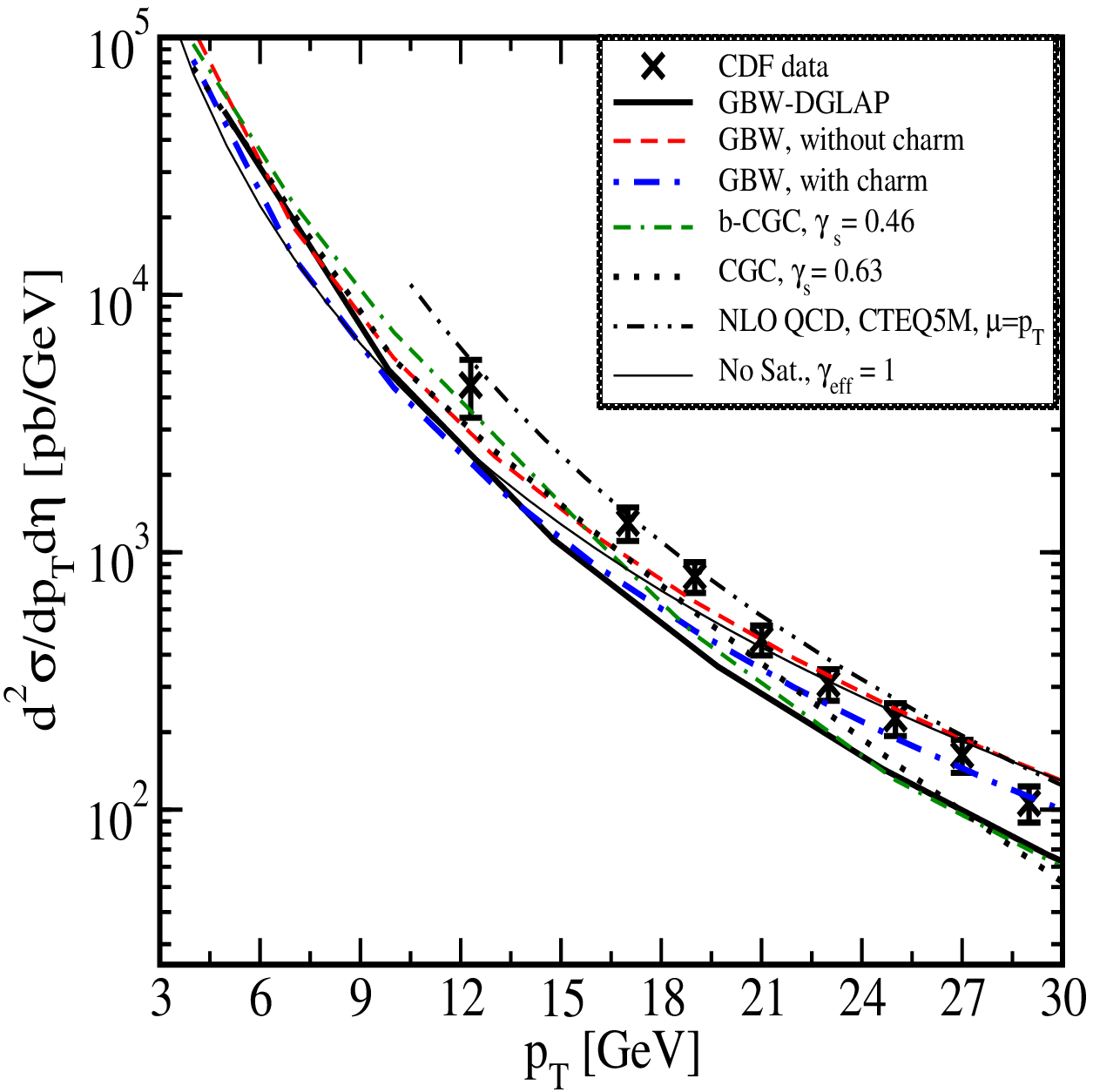}} \caption{ Inclusive direct photon spectra
calculated with various color-dipole models at mid-rapidity $\eta=0$
at the energies of RHIC, $\sqrt{s}=200$ GeV (upper panel), and
Tevatron, $\sqrt{s}=1.8$ TeV (lower panel).  The NLO QCD curve is from
the authors of reference \cite{ppqcd} (given in table 3 of
Ref.~\cite{cdf1}).  Experimental data are from the PHENIX experiment
\cite{rhic} at $\eta=0$, and from the CDF experiment \cite{cdf1,cdf} at
$|\eta|<0.9$.  The error bars are the linear sum of the statistical
and systematic uncertainties. \label{f2}}
\end{figure}

Notice that calculation of the $p_T$-distribution given by Eq.~(\ref{m1}) needs only knowledge of the total dipole cross section and is independent of the impact parameter dependence of the partial elastic dipole-proton amplitude. Nevertheless, we consider also the model proposed by Watt and Kowalski \cite{bccc}.
Although the main focus of this model is the impact parameter dependence (b-CGC), which is irrelevant for our calculations, the integrated cross section is different from the above mentioned models,

\begin{equation} \label{eq:bcgc}
  \sigma_{q\bar{q}}^{\text{b-CGC}}(x,r) = 2\int d^2b\,
  \sigma_{q\bar{q}}^{\text{CGC}}(x,r,b)\,,
\end{equation}
where $\sigma_{q\bar{q}}^{\text{CGC}}(x,r,b)$ is given by Eq.~(\ref{cgc}) with 
the saturation scale $Q_s$ which now depends on impact parameter,
\begin{equation} \label{bcgc1}
  Q_s\equiv Q_s(x,b)=\left(\frac{x_0}{x}\right)^{\frac{\lambda}{2}}\;\left[\exp\left(-\frac{b^2}{2B_{\rm CGC}}\right)\right]^{\frac{1}{2\gamma_s}}. 
\end{equation}
The parameter $B_{\rm CG}=7.5 \text{GeV}^{-2}$ is fitted to the
$t$-dependence of exclusive $J/\Psi$ photoproduction. It has been
shown that if one allows the parameter $\gamma_s$ to vary along side
with other parameters (in contrast with CGC fitting procedure where
$\gamma_s$ is fixed with LO BFKL value), it results in a significantly
better description of data for  $F_2$ with the value of
$\gamma_s=0.46$, which is remarkably close to the
value of $\gamma_s=0.44$ recently obtained from the BK equation \cite{bkk}.
Other parameters obtained from the fit are: $\mathcal N_0=0.558$,
$x_0=1.84\times 10^{-6}$ and $\lambda=0.119$.

 In order to demonstrate the importance of saturation, we will
 also use a non-saturated model (No Sat) fitted to $F_2$ with $x\le 0.01$ and 
$Q^2 \in[0.25, 45]~ \text{GeV}^2$: 
 \begin{equation} 
 \frac{d\sigma_{q\bar{q}}^{\text{No Sat}}}{d^2\vec{b}}=2\mathcal{N}_0\left(\frac{rQ_s}{2}\right)^{2\gamma_{eff}}  \label{nos}
\end{equation}
where $Q_s$ is defined in Eq.~(\ref{bcgc1}). The parameter $\gamma_{eff}$ is defined for $rQ_s\le 2$ as $\gamma_{eff}=\gamma_s +\frac{1}{\kappa\lambda Y}\ln\frac{2}{rQ_s}$,
and for $rQ_s> 2$ as $\gamma_{eff}=\gamma_s$. The other parameters are given by
$\gamma_s=0.43$, $\mathcal N_0=0.568$, $x_0=1.34\times 10^{-6}$ and
$\lambda=0.109$ \cite{bccc}.  Surprisingly, the fit obtained with such
an oversimplified model is as good as the other models with
$\chi^2/\text{d.o.f.}=0.92$ (although it should certainly fail to
explain data on diffractive DIS \cite{kps}, which are sensitive mainly
to large size dipoles). Notice that we use here the No Sat model for a
qualitative argumentation.

At first sight this result could be used as an argument that the data
is not sensitive to the saturation effect. However, the actual meaning
of this exercise is quite opposite. It is well known that the
saturation effects start being essential when the anomalous dimension
reaches the value $\gamma_{cr} = 1-\gamma_{eff}=0.37$ (see
Refs.~\cite{glr,GAMMA,MUT}). We will show that at very forward rapidities
at LHC, the diffusion term in the anomalous dimension is less important.
Therefore, what we actually demonstrate is that
the value of the anomalous dimension should be larger than
$\gamma_{cr}$ at very forward rapidities for LHC, and because of this the saturation effects have to be
taken into account.  

The second comment on this model (see Ref.~\cite{bccc}) is that it is 
actually a model which contains saturation, and the difference with
the CGC model (see \eq{cgc}) is only one: this model is written
for dipoles with sizes close to $1/Q_s$.  Indeed, comparing
\eq{cgc} and \eq{nos} one can see that they treat differently the
region $r\,Q_s \,>\,1$. The CGC model describes this region as
solution to the BK equation deeply in the saturation region \cite{LT},
with a phenomenological matching at $r\,Q_s = 2$, while this model uses
the solution to the BK equation \cite{MUT,IIM,LT} for $r\,Q_s >\,2$ but
$r$ close to $1/Q_s$. Therefore, it is not appropriate to call this
model ``no saturation model'', nevertheless, we use this name as a
terminology.

Summarizing, we can claim that direct photon production is sensitive
to saturation effects. In conclusion, the success of the so-called `no saturation
model' can be interpreted such that at the LHC we will be
still sensitive to the kinematic region close to the saturation
scale.

\begin{figure}[!t]
       \includegraphics[width=8 cm] {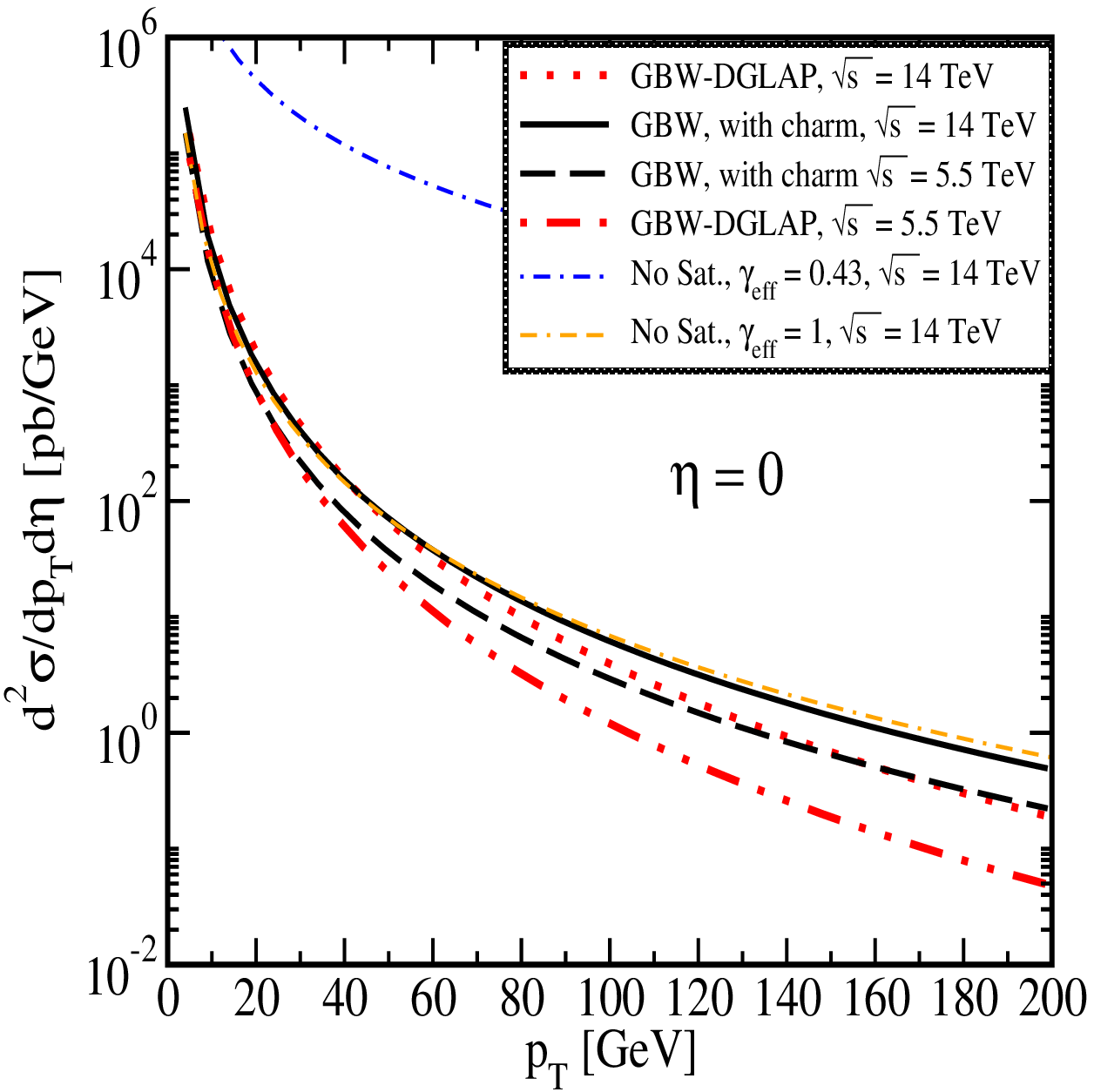}
       \includegraphics[width=8.1 cm] {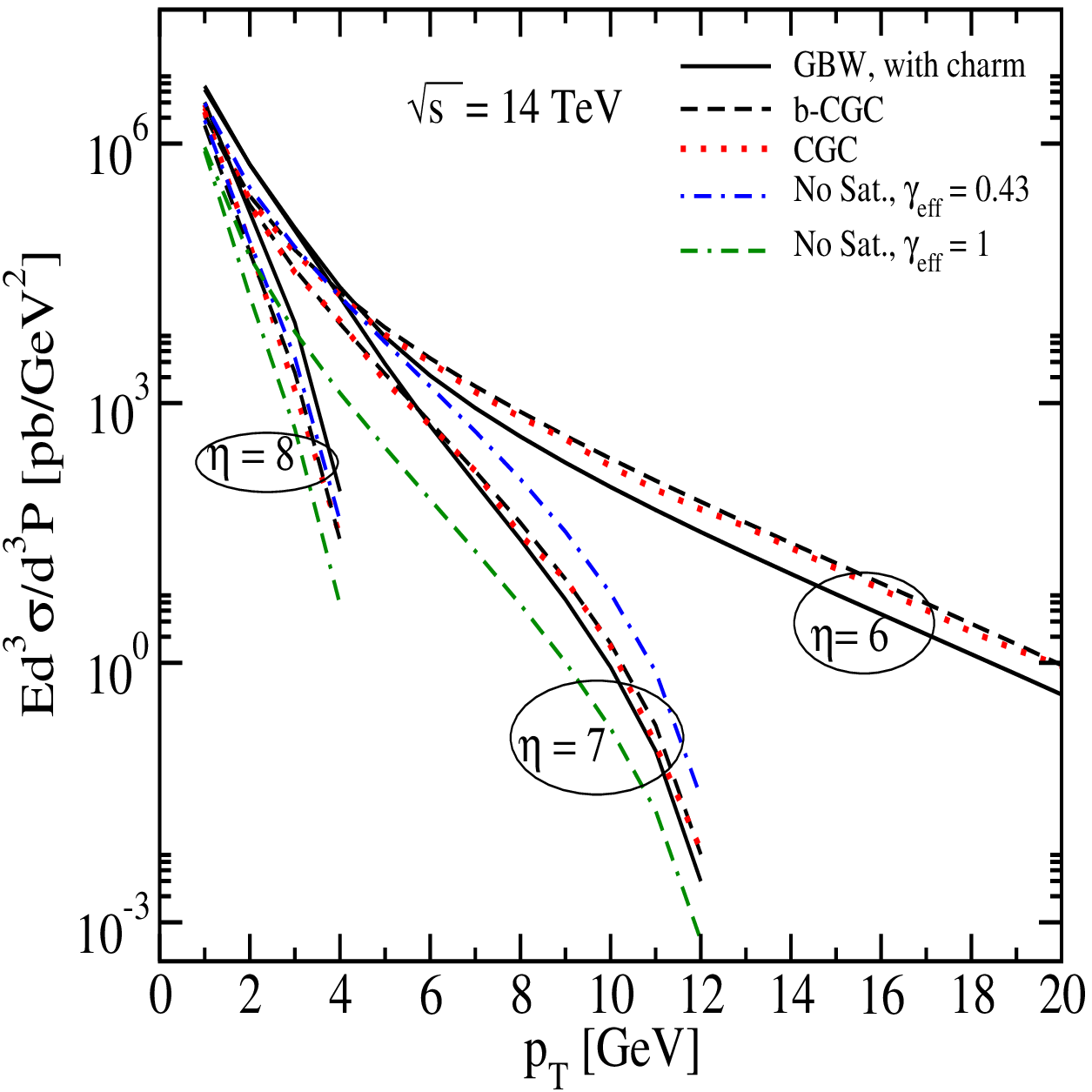}
       \caption{ Direct photon spectra obtained from various dipole model at
midrapidity (upper panel) and forward rapidities (lower panel) at the LHC energies for $pp$ collisions. \label{flhc}}
\end{figure}

\section{Numerical results and discussion}

In Fig.~\ref{f2}, we compare predictions of various dipole models with data for inclusive
prompt-photon production from RHIC at $\sqrt{s}=200$ GeV and from the Tevatron at
$\sqrt{s}=1.8$ TeV. A word of caution is in order here.
All the above parametrizations for the dipole cross-section have been fitted to DIS data at $x\le 0.01$. This corresponds to
$p_T \le 2$ GeV at the RHIC energy , so the PHENIX data plotted in the upper panel of Fig.~\ref{f2} are
not suited for a model test. Notice that the CDF data plotted in the bottom panel of Fig.~\ref{f2} were obtained with a so called isolation cut, which is aimed at suppression of the overwhelming background
of secondary photons originated from radiative hadron decays. This might change
the cross-section within $10-20\%$ for the CDF kinematics \cite{cdf}. 

One can see from Fig.~\ref{f2} that various dipole models presented in
the previous section with explicit saturation give rather similar
results at small $p_T$. At high $p_T$, CGC, b-CGC and GBW-DGLAP models
which incorporate QCD evolution provides a better description of data
compared to the GBW model. The No Sat model with the diffusion term, defined in
Eq.~(\ref{nos}), gives similar results as the b-CGC model for both
RHIC and CDF energies (not shown in the plot). In order to understand
the role of the diffusion term in the the anomalous dimension, we show
in Fig.~\ref{f2}, the results with two extreme limits
$\gamma_{eff}=0.43, 1$.  It is clear that the dipole model without
explicit saturation as given by Eq.~(\ref{nos}) with
$\gamma_{eff}=0.43$, does not describe the data either from PHENIX, or
from CDF (we do not show in Fig.~\ref{f2}, No Sat. with
$\gamma_{eff}=0.43$ curve for CDF, since it is about two orders of
magnitude above the other models and data). However, changing the
anomalous dimension to the DGLAP value with $\gamma_{eff}=1$,
dramatically changes the results and brings the curves (No Sat model)
at both energies of RHIC and Tevatron (at $\eta=0$) within the ranges
of other dipole models with saturation and the No Sat model in the presence of the diffusion term. Therefore, the diffusion term in the anomalous dimension
is very important at both RHIC and CDF energies.

\begin{figure}[!t]
                 \centerline{\includegraphics[width=8 cm]
                 {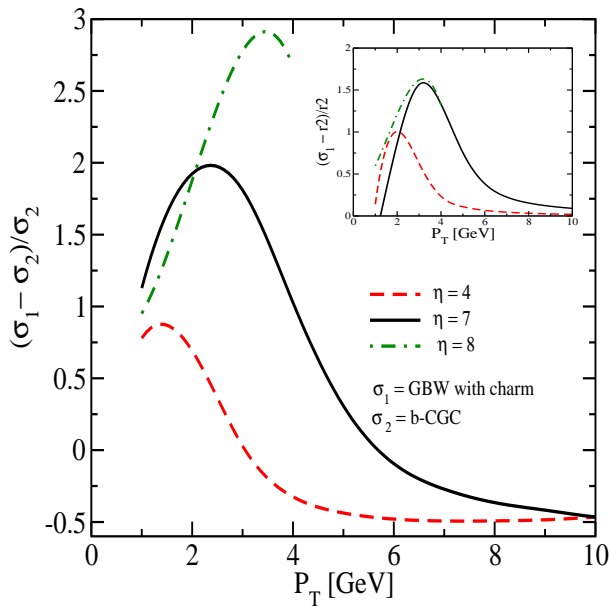}} \caption{The difference between the
                 GBW and the b-CGC models for the LHC energy $\sqrt{s}=14$
                 TeV at different rapidities. Inserted plot: the
                 discrepancy between the GBW and the $r2$ model (the
                 GBW model without explicit saturation) at various
                 rapidities for $\sqrt{s}=14$ TeV in $pp$ collisions.  \label{ftt}}
\end{figure}

We have recently shown that the
color dipole formulation coupled to the DGLAP evolution
provides a better description of data at large transverse
momentum compared to the GBW dipole model \cite{amir0}. In Fig.~\ref{flhc}, upper
panel, we show the predictions of the GBW (with charm quark) and the
GBW-DGLAP models for LHC energies $\sqrt{s}=5.5, 14$ GeV at
midrapidity for the transverse momentum up to $p_T=200$ GeV.  In Fig.~\ref{flhc}, lower panel, we show the predictions
of various color-dipole models for $\sqrt{s}=14$ GeV at different
rapidities. Generally, the discrepancy among predictions of various
models at moderate $p_T$ is not very large.  This can be also seen from
Fig.~\ref{ftt} where we compare, as an example, 
the GBW and b-CGC models, which expose very different structures (see Eqs.~(\ref{gbw},\ref{eq:bcgc}) and Fig.~(\ref{fd})). In the inserted plot
in Fig.~\ref{flhc}, we show the effect of unitarization within the
GBW model, namely using the exponent in Eq.~(\ref{gbw}) as a dipole
cross-section ($r2$ model). One can also see that the discrepancy between the GBW and the $r2$ model increases at forward rapidities, though it is still not appreciable.  

\begin{figure}[!]
       \centerline{\includegraphics[width=8.0 cm] {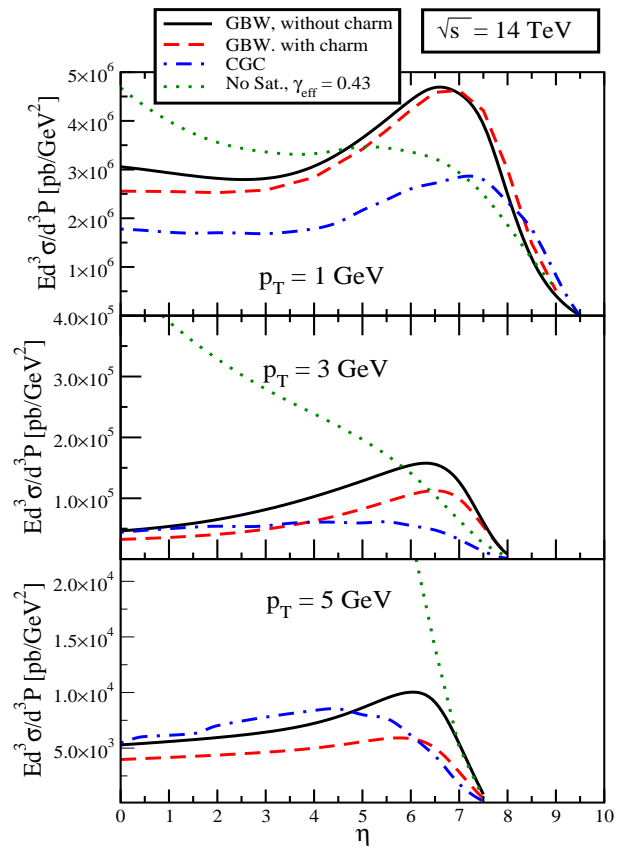}}
       \caption{ Invariant cross-section for direct photon production in $pp$ collisions as a function of rapidity calculated with various color
       dipole models
       $\eta$ for various fixed $p_T$ shown in the plots. \label{eta}}
\end{figure}

In Fig.~\ref{flhc}, we also show the results for the model without
saturation. At the midrapidity again the results of the No Sat
model with the DGLAP anomalous dimension $1-\gamma_{eff}=0$ is close
to other saturated models. At the same time, at very forward
rapidities, the anomalous dimension $1-\gamma_{eff}=0.57$ which is
close to the value predicted from the BK equation \cite{bkk}, will be
in favour of other models. At forward rapidities, the diffusion term in 
the anomalous dimension is not important more, since it gives similar result as with a
fixed $\gamma_{eff}=0.43$. This indicates that direct photons
production at different rapidities at LHC is rather sensitive to the
saturation.  Again, since the values of anomalous dimension turn out to
be larger than $\gamma_{cr}=0.37$, such a description of the
experimental data indicates at a large saturation effect.

In Fig.~\ref{eta}, the differential cross-section of photon radiation at the energy of LHC is plotted versus rapidity  at  fixed
transverse momenta $p_T=1,3, 5$ GeV. Calculations were performed with several models for the dipole  cross section. 
All of them lead to a substantial enhancement of the photon production rate  at forward
rapidities.  One can see that the
larger the saturation scale is, the stronger is the peak. From
Fig.~\ref{eta}, it is again obvious that No Sat model at very forward
rapidity is within the dipole model`s predictions with an explicit
saturation. However, the peak disappeared since at about mid-rapidity
$\gamma_{eff}=0.43$ is too small. In principle, the
peak can be also reproduced in the No Sat model if one allows an
anomalous dimension running  with energy and transverse momentum $p_T$.  The appearance of
the peak at forward rapidity is a direct consequence of the abelian nature of the
electromagnetic interaction. In the case of gluon radiation, the peaks at
forward-backward rapidities is be replaced with a kind of plateau at central rapidities, which is indeed observed in data for hadron production.

To conclude, by this letter we encourage measurements of direct
photons at forward rapidities in $pp$ collisions at modern colliders
. These experiments will be a sensitive tool for search for saturation
effects, since they will allow to access the smallest possible values
of Bjorken $x$ in the target. Besides, the background of photons from
radiative hadronic decays should be significantly suppressed. As we
demonstrate in Fig.~\ref{eta}, direct photons are enhanced, even form
a bump, at forward rapidities. At the same time, gluon nonabelian
radiation is known to be strongly suppressed in this region,
 so hadron and decay photons are also suppressed. We provided predictions for the cross section of direct photon
production at various rapidities for $pp$ collisions at LHC employing
different models for the dipole-proton total cross section.

 \begin{acknowledgments} 
We would like to thank Graeme Watt for pointing out typos and useful communication. 
This work was supported in part by Conicyt (Chile) Programa Bicentenario PSD-91-2006, by Fondecyt (Chile) grants 1070517, 1050589 and by DFG (Germany) grant PI182/3-1, by BSF grant $\#$ 20004019, by a grant from Israel Ministry of Science, Culture \& Sport, and
the Foundation for Basic Research of the Russian Federation.

\end{acknowledgments}

\end{document}